\documentclass[12pt,preprint]{aastex}
\usepackage{amsmath,bm}
\usepackage{multirow}
\usepackage{enumerate}
\usepackage{graphicx}
\usepackage{color}
\usepackage[T1]{fontenc}
\usepackage{url}
\usepackage{hyperref}
\newcount\doepsf
\newcommand{\be}{\begin{equation}}
\newcommand{\ee}{\end{equation}}
\newcommand{\bea}{\begin{eqnarray}}
\newcommand{\eea}{\end{eqnarray}}
\newcommand{\bean}{\begin{eqnarray*}}
\newcommand{\eean}{\end{eqnarray*}}

 \definecolor{DarkGreen}{rgb}{0.0,0.45,0.0}     
 \definecolor{DarkMagenta}{rgb}{0.45,0.0,0.45}  

\begin{document}

\title{Onset of secondary instabilities and plasma heating during magnetic reconnection in strongly magnetized regions of the low solar atmosphere }

\author{Lei Ni$^{1, 2, 3}$,
             and Vyacheslav S. Lukin$^{4}$\footnote{Any opinion, findings, and conclusions or recommendations expressed in this material are those of the authors and do not necessarily reflect the views of the National Science Foundation} ,
            }

\affil{$^1$Yunnan Observatories, Chinese Academy of Sciences, P. O. Box 110, Kunming, Yunnan 650216, P. R. China}
\affil{$^2$CAS Key Laboratory of Solar Activity, National Astronomical Observatories, Beijing 100012, P. R. China}
\affil{$^3$Center for Astronomical Mega-Science, Chinese Academy of Sciences, 20A Datun Road, Chaoyang District, Beijing 100012, P. R. China}
\affil{$^4$National Science Foundation, 2415 Eisenhower Ave, Alexandria, VA 22314, USA}

\shorttitle{Plasmoid instabilities around the temperature minimum region}  
\shortauthors{Ni et al.}

\email{leini@ynao.ac.cn}

\slugcomment{MR in TMR; v.\ \today}

\begin{abstract}
\noindent  We numerically study magnetic reconnection on different spatial scales and at different heights in the weakly ionized plasma of the low solar atmosphere (around $300-800$~km above the solar surface) within a reactive 2.5 D multi-fluid plasma-neutral model. We consider a strongly magnetized plasma ($\beta \sim 6\% $) evolving from a force-free magnetic configuration and perturbed to initialize formation of a reconnection current sheet. On large scales, the resulting current sheets are observed to undergo a secondary 'plasmoid' instability. A series of simulations at different scales demonstrate a cascading current sheet formation process that terminates for current sheets with width of ~2m and length of $\sim100$~m, corresponding to the critical current sheet aspect ratio of $\sim50$. We also observe that the plasmoid instability is the primary physical mechanism accelerating the magnetic reconnection in this plasma parameter regime. After plasmoid instabilities appear, the reconnection rate sharply increases to a value of $\sim$ 0.035, observed to be independent of the Lundquist number. These characteristics are very similar to magnetic reconnection in fully ionized plasmas. In this low $\beta$ guide field reconnection regime, both the recombination and collisionless effects are observed to have a small contribution to the reconnection rate. The simulations show that it is difficult to heat the dense weakly ionized photospheric plasmas to above $2\times10^4$~K during the magnetic reconnection process. However, the plasmas in the low solar chromosphere can be heated above $3\times10^4$~K with reconnection magnetic fields of $500$~G or stronger.    
\end{abstract}   


\section{Introduction}
\label{s:introduction}
Magnetic reconnection is a fundamental magnetized plasma process in the heliosphere. In the partially ionized low solar atmosphere, numerous small-scale magnetic reconnection events have been observed by both the ground and space based high resolution Solar telescopes. Examples of  such observations include microflares \citep[e.g.,][]{1999ApJ...512..454D, 2009ApJ...692..492B}, chromospheric jets \citep[e.g.,][]{2009ApJ...707L..37L, 2013A&A...552L...1B, 2014Sci...346A.315T, 2017MNRAS.464.1753H}, type II spicules \citep[e.g.,][]{2000SoPh..196...79S, 2012ApJ...752..108S}, Ellerman bombs (EBs) \citep[e.g.,][]{2006ApJ...643.1325F, 2014ApJ...792...13H, 2015ApJ...798...19N} and the high temperature IRIS bursts (IBs) \citep[e.g.,][]{2014Sci...346C.315P, 2015ApJ...812...11V, 2016A&A...593A..32G, 2016ApJ...824...96T, 2017ApJ...851L...6R, 2018ApJ...854..174T}. At present, there are at least two challenges for a magnetic reconnection model in the low solar atmosphere: the plasma heating and the reconnection rate in a viable model should match the observational data. 

One of the most puzzling problems of the low solar atmosphere reconnection events is whether or not the IBs are related to the traditional EBs. The traditional EBs are usually observed in the upper solar photosphere or the low chromosphere. A semi-empirical model of a variety of combinations of the spectral lines originally led to estimates of the temperature increases in EBs ranging from $\sim400-2000$~K \citep[e.g.,][]{2006ApJ...643.1325F, 2014A&A...567A.110B, 2016A&A...593A..32G}. The IBs observed by the IRIS Si IV 'Transition Region' (TR) lines imply that the temperature in IBs can be above $8\times10^4$~K \citep{2014SoPh..289.2733D, 2014Sci...346C.315P}. \cite{2014Sci...346C.315P} suggested that such IBs could be evidence of heating within photospheric EBs with temperatures an order of magnitude higher than those predicted by previous semi-empirical models. The recent joint observations from the ground and space based telescopes also support the assertion that the plasmas within some of the photospheric active region EBs and quiet Sun Ellerman-like brightenings (QSEBs) can be strongly heated to reach TR temperatures \citep[e.g.,][]{2015ApJ...812...11V, 2016ApJ...824...96T, 2017ApJ...845...16N, 2018ApJ...854...92T}. However, \cite{2016A&A...590A.124R} suggested that temperatures as low as $2\times10^4$~K could account for the observed increased emission in the Si IV line. In their work, they assumed the local thermodynamic equilibrium (LTE) for line extinctions during the onsets of EBs.

Several previous single-fluid MHD simulations showed that the maximum temperature increases in magnetic reconnection regions below the upper chromosphere are always only several thousand K \citep[e.g.,][]{2007ApJ...657L..53I, 2009A&A...508.1469A, 2011RAA....11..225X, 2017ApJ...839...22H}. However, the resolution in these simulations was not sufficient to resolve the current sheet sub-structure and the artificial hyper-diffusivity operator that was included to prevent the collapse of the current sheets leaves open the possibility of formation of hotter smaller scale. The high resolution simulations with the more realistic magnetic diffusion in \cite{2016ApJ...832..195N} showed that the plasma can be heated from $4200$~K to above $8 \times10^4$~K inside the multiple magnetic islands of a reconnection process with strong magnetic fields ($500$~G) in the temperature minimum region (TMR) of the solar atmosphere. Ambipolar diffusion, temperature-dependent magnetic diffusion, heat conduction, and optically thin radiative cooling were all included in \cite{2016ApJ...832..195N}. However, some of the important interactions between ions and neutrals were ignored and the non-equilibrium ionization-recombination effects were not considered in any of the single-fluid MHD simulations. The recent reactive multi-fluid plasma-neutral simulations \citep{2018ApJ...852...95N, 2018PhPl...25d2903N} with HiFi code showed that the plasmas in TMR can be heated above $2\times10^4$~K when the reconnection magnetic field strength was greater than $500$~G. \cite{2018ApJ...852...95N} has also proved that the non-equilibrium ionization-recombination dynamics play a critical role in determining the structure of the reconnection region, and it leads to much lower temperature increases as compared to simulations that assume plasma to be in ionization-recombination equilibrium. The high temperature increases in the single-fluid simulations might be overestimated.     	

The spatial scales, lifetimes, plasma flow velocities and emission intensity of the low solar atmospheric reconnection events are usually much smaller than the eruptions in the solar corona.  According to the length scale, lifetime and the estimated Alfv\'{e}n speed (or the reconnection outflow speed), the reconnection rate in the low solar atmosphere events can be estimated as $M \approx L/(V_{A} t_{life}) \approx L/(V_{out} t_{life})$, which is about $0.01-0.5$\citep[e.g.,][]{2011ApJ...731...43N, 2012ApJ...760..109L, 2015ApJ...799...79N}. The magnetic diffusion coefficient including both the electron-ion and electron-neutral collisions \citep{2012ApJ...760..109L} is estimated order of $10^3-10^4$ m$^2$\,s$^{-1}$ from the upper photosphere to the middle chromosphere. The corresponding Lundquist number (magnetic Reynolds number) for a length scale $L\sim1$ Mm and Alfv\'{e}n speed $V_A\sim(10-100)$ km\,s$^{-1}$ is of order $S=LV_A/\eta \sim 10^6-10^8$. Thus, the reconnection rate for the classical Sweet-Parker current sheet is calculated as $M_{SP}\sim S^{-1/2} \sim 10^{-4}-10^{-3}$ from the one-fluid model, which is far smaller than the reconnection rate required in the low solar atmosphere reconnection events.  

The plasmas around the TMR are weakly ionized and the plasma density is very high. Thus, it is important to include the interactions between ions and neutral particles as well as the radiative cooling in any realistic model of magnetic reconnection around the TMR. The previous 1D analytical work \citep[e.g.,][]{1999ApJ...511..193V, 2003ApJ...583..229H, 2004ApJ...603..180L} found that the high recombination can produce a loss of ions in the reconnection region that prevents ion pressure building up, which leads to faster magnetic reconnection independent of magnetic diffusivity. \cite{2012ApJ...760..109L, 2013PhPl...20f1202L} used the reactive multi-fluid plasma-neutral module within the HiFi modeling framework to study null-point magnetic reconnection in the solar chromosphere. They showed that even for the Lundquist number as high as $10^5$, the reconnection rates can reach above $0.05$, which is much faster than that predicted from the one-fluid Sweet-Parker model. They concluded that the strong ion recombination in the reconnection region, combined with Alfv\'{e}nic outflows,  can lead to fast null-point reconnection within a two-dimensional (2D) numerical model. The plasmoid instability only increased the reconnection rate by approximately $15\%$ and the Hall effects were negligible in their simulations. The recent 2.5 D simulations by \cite{2018ApJ...852...95N, 2018PhPl...25d2903N} used the same reactive multi-fluid plasma-neutral module within the HiFi modeling framework to model guide-field, low $\beta$  reconnection within a weakly ionized plasma. Their results showed that the decoupling of the ionized and neutral fluids was not obvious in the strongly magnetized regions with much lower plasma $\beta$, and the ionization rate was much faster than the recombination rate, which was very different from the previous high $\beta$ null-point simulations by \cite{2012ApJ...760..109L, 2013PhPl...20f1202L}. Though the strong radiative cooling was shown to result in loss of  most of the generated thermal energy, it also did not noticeably accelerate magnetic reconnection \citep{2018ApJ...852...95N, 2018PhPl...25d2903N}, which is different from the high $\beta$ simulations with radiative cooling in \cite{2017ApJ...842..117A}. 

In this work, we present 2.5 D simulations of magnetic reconnection on different spatial scales and at different heights above the solar surface by continuing to use  the reactive multi-fluid plasma-neutral module within the HiFi modeling framework. Comparing with our previous work \citep{2018ApJ...852...95N, 2018PhPl...25d2903N}, the simulations are performed on much larger scales. We find the critical Lundquist number and current sheet aspect ratio for the onset of the plasmoid instability in an initially strongly magnetized, weakly ionized plasma in the solar TMR. The physical mechanism that leads to fast magnetic reconnection in the higher Lundquist number reconnection process in the strongly magnetized solar TMR is discussed.  Section II describes the numerical model and initial conditions. The numerical results are presented in Section III. A summary and discussion are given in Section IV.

\section{Numerical model and initial conditions}
\label{s:model}
The same reactive multi-fluid plasma-neutral module of the HiFi modeling framework as used by \cite{2018ApJ...852...95N} is applied in this work. We still normalize the equations by using the characteristic plasma density and magnetic field around the strongly magnetized solar TMR. The characteristic plasma number density is $n_{\star}=10^{21}$~m$^{-3}$, and the characteristic magnetic field is $B_{\star}=0.05$~T=$500$~G. From these quantities, we derive additional normalizing values to be $V_{\star}\equiv B_{\star}/\sqrt{\mu_0m_pn_{\star}}=34.613$~km\,s$^{-1}$, $T_{\star} \equiv B_{\star}^2/(k_B\mu_0n_{\star})=1.441\times10^5$~K. There are three characteristic length scales in our simulations: $L_{\star 1}=200$~m, $L_{\star 2}=1000$~m and $L_{\star 3}=10^4$~m. The three corresponding characteristic times are $t_{\star 1}\equiv L_{\star 1}/V_{\star}=0.0058$~s, $t_{\star 2}\equiv L_{\star 2}/V_{\star}=0.029$~s and $t_{\star 3}\equiv L_{\star 3}/V_{\star}=0.29$~s, respectively. 

The simulations are initialized with a force-free Harris sheet magnetic equilibrium, where the reconnecting magnetic field is along the $y$-direction and the guide B-field is in the $z$-direction. Specifically, the initial dimensionless magnetic flux in the $z$-direction is given by 
\begin{equation}
  A_{z0}(y)=-b_p\lambda_{\psi} \mathrm{ln} \left[\mathrm{cosh}\left(\frac{y}{\lambda_{\psi}}\right)\right],
\end{equation}
where $b_p$ is the strength of the the magnetic field and $\lambda_{\psi}$ is the initial thickness of the current sheet. The initial magnetic field in z-direction is  
\begin{equation}
  B_{z0}(y)=b_p \bigg/  \left[ \mathrm{cosh}\left(\frac{y}{\lambda_{\psi}}\right) \right].
\end{equation}

Since the ion inertial length is much smaller than the width of the current sheets in this work, the Hall effect is not important and ignored. The numerical results in our previous paper \cite{2018ApJ...852...95N} demonstrate that the collisions between electrons and neutrals are not important for magnetic reconnection in low $\beta$ plasmas around the solar TMR. We also ignore the collisions between electrons and neutral particles, and the dimensionless magnetic diffusivity in this work is 
\begin{equation}
   \eta=\eta_{ei}=\eta_{ei\star}T_e^{-1.5},
\end{equation}  
where $\eta_{ei\star}$ is a normalization constant derived from the characteristic values $n_{\star}$, $B_{\star}$, and $L_\star$. One can get $\eta_{ei\star 1}=3.729\times10^{-6}$, $\eta_{ei\star 2}=7.457\times10^{-7}$ and $\eta_{ei\star 3}=7.457\times10^{-8}$ by using the three different length scales. $T_e$ is the dimensionless electron temperature. The electron and ion fluids are assumed to be well-coupled and only the hydrogen gas is considered in our model. Thus, we assume $T_i=T_e$, $n_i=n_e$, and the pressure of the ionized component is twice the ion (or electron) pressure, $P_p = P_e + P_i = 2P_i$.

In this work, we have separately tested our simulations by using two different radiative cooling functions in the plasma internal energy equation as presented in our previous papers \citep{2018ApJ...852...95N, 2018PhPl...25d2903N}. The first simple radiative cooling model \citep{2011PhDthesis, 2012ApJ...760..109L} represents the radiative losses that are due to atomic physics such as radiative recombination, and the excited states are not tracked. The formula of this simple model is given by
\begin{equation}
  L_{rad1} = \Gamma_i^{ion}\phi_{eff},
\end{equation}
where $\phi_{eff}=33$~eV$=5.28\times10^{-18}$~J. The ionization rate $\Gamma_i^{ion}$ is defined as
\begin{equation}
   \Gamma_i^{ion}=  \frac{n_n n_e A}{X+\phi_{\mathrm{ion}}/T_e^{\ast}}\left(\frac{\phi_{\mathrm{ion}}}{T_e^{\ast}}\right)^{K} \mathrm{exp}(-\frac{\phi_{\mathrm{ion}}}{T_e^{\ast}}), 
\end{equation}
using the values $A=2.91\times10^{-14}$, $K=0.39$, $X=0.232$, and the hydrogen ionization potential $\phi_{\mathrm{ion}}=13.6$~eV. The unit for the neutral and electron number density are both m$^{-3}$, $T_e^{\ast}$ is the electron temperature specified in eV. Then we can get the unit for $\Gamma_i^{ion}$ is m$^{-3}$\,s$^{-1}$ and the unit for $L_{rad}$ is J\,m$^{-3}$\,s$^{-1}$. The radiative cooling goes to zero if the plasma is fully ionized. As shown in our previous work \citep{2018ApJ...852...95N}, the plasmas will be fully ionized if the reconnection magnetic fields are strong enough. This simple radiative cooling model becomes invalid in this situation. The second radiative cooling function is computed by using the OPACITY project and the CHIANTI databases \citep{2012ApJ...751...75G}. This radiative cooling model is considered to be a more realistic cooling model for plasmas in the solar chromosphere \citep{2017ApJ...842..117A}. The expression of this radiative model in units with J\, m$^{-3}$\,s$^{-1}$ is 
\begin{equation}
  L_{rad2} = C_E n_e (n_n+n_i) 8.63\times10^{-6}T^{-1/2} \times \sum_{i=1}^2 E_i \Upsilon_i exp(-eE_i/k_BT), 
\end{equation}
where $C_E=1.6022\times10^{-25}$~J\,eV$^{-1}$, $E_1=3.54$~eV and $E_2=8.28$~eV, $\Upsilon_1=0.15\times10^{-3}$ and $\Upsilon_2=0.065$. A three level hydrogen atom with two excited levels is included in this function, and $E_1$ and $E_2$ are the excited level energies. The unit for the temperature in this equation is Kelvin and the unit for number density is m$^{-3}$. The background constant heating is included to balance the initial radiative cooling when the second radiative cooling model is applied. The heating function $H_{0}$ is equal to the initial value of $L_{rad2}$. In our simulations, we have normalized $L_{rad1}$ in equation(4) and $L_{rad2}$ in equation (6) by using the characteristic values presented above.  

The two radiative cooling models with the remaining simulation setup being identical for each case have been tested, we find that the main conclusions in this work do not dependent on the two different radiative cooling terms applied in the code. We will present the numerical results of five cases in this work, Case~I, II, III, IV and V. In order to compare with Case~A in our previous work \citep{2018ApJ...852...95N}, Case~I, II, III and IV show the results with the simple radiative cooling model $L_{rad1}$. Since the plasmas in Case~V will be fully ionized if the simple radiative cooling model is applied, Case~V shows the results with the more realistic radiative cooling model $L_{rad2}$. Both a higher resolution run and a lower resolution run have been carried out  for each case.

Case~I, II and III represent magnetic reconnection around the solar TMR ($\sim600$~km above the solar surface) on three different scales. All the initial parameters in Case I, II and III are the same, except for the simulation length scales. The characteristic length scales in Cases I, II and III are $L_{\star 1}$, $L_{\star 2}$ and $L_{\star 3}$ as presented above, respectively. The initial neutral particle number density and ionization fraction in these three cases are $n_{n0}=0.5n_{\star}$ and  $f_{i0}=0.01\%$. Since the hydrogen gas only has the ground state and the ionized state in this model, the initial temperatures of the ionized and neutral fluids are set to be uniform at $T_{i0}=T_{n0}=0.05829T_{\star}$ in Case~I, II and III to keep the initial ionization fraction $f_{i0}=0.01\%$, which is the same as that around the solar TMR. The strength of the initial magnetic field in dimensionless form is $b_p=1$ in Case I, II and III.  

Case~IV represents magnetic reconnection in the photosphere (around $300$~km above the solar surface) with a length scale of $L_{\star 1}=200$~m. The initial neutral particle number density and ionization fraction in Case~IV are $n_{n0}=12.5n_{\star}$ and  $f_{i0}=0.01\%$. The initial temperatures of the ionized and neutral fluids are also set to be uniform at $T_{i0}=T_{n0}=0.05829T_{\star}$ in this case. The initial magnetic field in dimensionless form in Case~IV is $b_p=5$, which is five times higher than that in Case~I. 

Case~V represents magnetic reconnection in the low solar chromosphere (around $800$~km above the solar surface) with a length scale of $L_{\star 1}=200$~m. The initial neutral particle number density and ionization fraction in Case~V are $n_{n0}=0.1n_{\star}$ and  $f_{i0}=0.1\%$. The initial temperatures of the ionized and neutral fluids are also set to be uniform at $T_{i0}=T_{n0}=0.06559T_{\star}$ in this case. The initial magnetic field in dimensionless form in Case~V is $b_p=1$, which is the same as in Case~I. 

The initial parameters in the five cases are presented in Table~1. We only simulate one quarter of the domain ($0<x<2L_{\star}$, $0<y<L_{\star}$) in all five cases, $L_{\star}$ are corresponding to $L_{\star 1}$, $L_{\star 2}$ and $L_{\star 3}$ in Case I, II and III, respectively. We use the same outer boundary conditions at $|y|=L_{\star}$ and the same form of the initial electric field perturbations as \cite{2015ApJ...805..134M} to initiate magnetic reconnection in all of the cases in this work. The perturbation electric field is applied for $0\leq t \leq t_{\star}$. The perturbation magnitude is proportional to the value of $b_p$ in each of the cases with the amplitude of $\delta E=10^{-3}b_p$. Periodicity of the physical system is imposed in the x-direction at $|x|=2L_{\star}$. Presented simulation results from all five cases have been tested by using a lower resolution and a higher resolution. The highest number of grid elements in Case III is $m_x=288$ elements in the $x$-direction and $m_y=240$ elements in the y-direction. We use sixth order basis functions for all simulations, resulting in effective total grid size $(M_x, M_y)=6(m_x,m_y)$. Grid packing is used to concentrate mesh in the reconnection region. Therefore, the mesh packing along the $y$-direction is concentrated to a thin region near $y=0$. The smallest normalized grid sizes along the $y$-direction near $y=0$ in both the higher and lower resolution runs of Case~I, II, III, IV and V are presented in Table~2.  

\section{Numerical Results}\label{s:results}
\subsection{Magnetic reconnection on different length scales}\label{ss:MR:I}
 In our previous papers \citep{2018ApJ...852...95N, 2018PhPl...25d2903N}, magnetic reconnection on a very small scale ($L_{\star}=100$\,m) with low Lundquist number ($\sim 2000$) has been studied. The plasmoid instability did not appear in those simulations. In this section, we will show the studies of magnetic reconnection on three larger scales: Case~I with $L_{\star1}=200$\,m, Case~II with $L_{\star2}=1000$\,m and Case~III with $L_{\star3}=10^4$\,m. The plasmoid instability appears in all of the three cases. Case~A in \cite{2018ApJ...852...95N} is used to compare with the three cases in this work. The only difference among the four cases (Cases~A, I, II and III ) is the characteristic length scale. In this work, the half length $L_{sim}$ and half width $\delta_{sim}$ of the current sheet are defined by using the half-width at half-maximum and half-length at half-maximum in the current density $J_{z}/J_{\star}$. The method for calculating the reconnection rate $M_{sim}$ in this work is the same as that in the previous papers \citep{2013PhPl...20f1202L, 2018ApJ...852...95N, 2018PhPl...25d2903N}, $M_{sim}=\eta^{\ast} j_{max}/(V_A^{\ast} B_{up})$, where $j_{max}$ is the maximum value of the out of plane current density $J_{z}/J_{\star}$ at the reconnection X-point $(x_{R}, y_{R})$. $B_{up}$ is $B_x$ evaluated at $(x_{R}, \delta_{sim})$. $V_{A}^{\ast}$ is the relevant  Alfv\'en velocity defined using $B_{up}$ and the total number density $n^{\ast}$ at the location of $j_{max}$. $\eta^{\ast}$ is the magnetic diffusion coefficient defined by Equation (3) at the location of $j_{max}$. The calculated Lundquist number $S_{sim}$ is defined as $S_{sim}=V_A^{\ast}L_{sim}/\eta^{\ast}$.    
 
 Figure.~1 presents the distributions of current density and ionization fraction during the later stage of the reconnection process in Cases~A, I, II and III. This figure reveals the cascading process of the plasmoid instability. Several magnetic islands appear inside the current sheet in Case~III. The outflow region is very turbulent in Case~III, many small scale current sheet fragments and small magnetic islands even appear in the outflow region. The small scale current sheet fragments also appear in the outflow regions of Case II and I, but this region is relatively smoother in the case with smaller length scale. We only identify one magnetic island in Case~I. There is no magnetic island inside the current sheet of  Case~A. The outflow region of Case~A is also very smooth, no current sheet fragments appear there. Therefore, the plasmoid instabilities terminate at the reconnection current sheet length of $100$~m.
 
Figure.~2(a) and 2(b) show the calculated aspect ratio of current sheets (CSs) $L_{sim}/\delta_{sim}$ and Lundquist number $S_{sim}$ versus time before magnetic islands appear in Cases A, I, II and III, both the higher and lower resolution results of Case~I,  II and III are presented. One should note that the characteristic length scale $L_{\star}$ and time $t_{\star}$ are different in the different cases. $t_{p1}$ is the dimensionless time point when the first magnetic island appears in each run. For the same case, one can find that magnetic islands appear later in the run with the higher resolution. The reason is that the higher resolution results in the lower level of numerical noises, which allows the Lundquist number to reach higher values before the plasmoid instability enters the nonlinear regime. The sensitivity of plasmoid-unstable high Lunquist number current sheets to noise is well known in the single-fluid MHD regime, e.g. see \cite{2010PhPl...17f2104H}, and is apparent here for Cases II and III. In Case~I, the aspect ratio reaches about $70$ and the Lundquist number is about $3400$ just before the first magnetic island appears in both the higher and the lower resolution runs. The aspect ratio reaches $45$ and the Lundquist number is about  $2000$ at the end of the simulation in Case~A, but no magnetic island appears in this case. Therefore, we can conclude that the critical aspect ratio is between $45-70$ and the critical Lundquist number is between $2000-3400$ for the plasmoid instability to appear in our simulations. These results are different from the previous work by \cite{2012ApJ...760..109L}. In their higher $\beta$ simulations with a 2D Harris sheet as initial magnetic field configurations, they find that the critical aspect ratio is about $200$ and the critical Lundquist number is about $10^4$ for the plasmoid instability. They point out that the magnetic islands which form due to the plasmoid instability in their simulations are losing ions due to recombination, and so their evolution is potentially very different from those in fully ionized simulations. In their later work with the more realistic magnetic diffusion \citep{2013PhPl...20f1202L}, they find that the plasmoid instability does not appear even when the aspect ratio is greater than $500$, which they argue may be due to the lower magnetic Prandtl number used in \cite{2013PhPl...20f1202L}, allowing for stronger shearing of the reconnection outflow and leading to stabilization of the plasmoid instability. However, we should point out that the critical Lundquist number and aspect ratio depend on many different factors, e.g. the amplitudes of viscosity and perturbation noises \citep{2016PhPl...23c2111C, 2016PhPl...23j0702C, 2017ApJ...850..142C, 2017ApJ...849...75H}; and the plasma $\beta$ and the initial current sheet configuration  \citep{2012PhPl...19g2902N, 2013PhPl...20f1206N} can impact the onset of plasmoid instability. In previous work on magnetic reconnection in fully ionized plasmas \citep{2012PhPl...19g2902N, 2013PhPl...20f1206N}, it was found that the critical Lundquist number in the 2D case with an initial Harris current sheet and nonuniform plasma pressure is much higher than that in the 2.5D case with  an initial force-free current sheet and uniform plasma pressure. The results presented here are consistent with these prior findings.


Prior work by \cite{2017ApJ...849...75H} indicates that the aspect ratio scales with the Lundquist number as $L_{sim}/\delta_{sim} \sim S^{1/2}$ for the lower Lundquist cases ($S < 2\times10^5$), and this scaling is closer to $L_{sim}/\delta_{sim} \sim S^{1/3}$ for the higher Lundquist number cases.  From Figure~2(a) and 2(b) in this work, we can find that the aspect ratios are $70$, $116$ and $120$ and the corresponding Lundquist numbers are $3400$, $11000$ and $17000$ at $t_{p1}$ just before the first magnetic island appears in Cases~I, II and III, respectively. The scaling in this work is closer to $L_{sim}/\delta_{sim} \sim S^{1/2}$. Therefore, our numerical results are consistent with \cite{2017ApJ...849...75H} .         

Figure~3(a) and (b) show the half width of the current sheet and reconnection rate versus time before and after plasmoid instabilities appear, both the higher and lower resolution results of Case~I,  II and III are presented. In both the higher and lower resolution runs, one can find that the half widths of current sheets (CSs) drop to about $2$~m, $5$~m and $20$~m at $t_{p1}$ just before the first magnetic island appears inside the current sheet in Cases~I, II and III, respectively. $t_{p2}$ is the time point when the second magnetic island begins to appear inside the current sheet in the lower resolution run of Case~III. After plasmoid instabilities appear, the current sheet width around the main reconnection X-point continues to drop sharply in all of the three cases. In both the higher and lower resolution runs, the thinnest half width eventually reaches about $1$~m and $2$~m in Case~I and II, respectively. It reaches about $7$~m in the lower resolution run of Case~III. One should note that the higher resolution simulation of Case~III is terminated when the plasmoid instability just develops into the nonlinear stage and the sharp decrease of the current sheet width does not happen yet. The resolution in the lower resolution run of  Case~III is obviously not high enough for the half width of the current sheet to reach $1$~m scale. In Case~I, the result in the higher resolution run confirms that the half width of the current sheet thins to 1 m without undergoing the higher order plasmoid instability. When the half width of the current sheet thins to $1$~m in Case~A, the plasmoid instability does not appear. Therefore, we can conclude that the plasmoid instability terminates at the reconnection current sheet width of $2$~m. 

From Figure~3(b), one can find that the reconnection rate in Case~III is the lowest in all of the three cases before the plasmoid instabilities appear, with the reconnection rate approximately scaling with the Lundquist number as $M_{sim} \sim S_{sim}^{-1/2}$ at $t_{p1}$. After plasmoid instabilities appear, the reconnection rates increase sharply in all of the three cases. The first peak of the reconnection rate in the lower resolution run of Case~III is about 0.015, which is about two times higher than that before the first magnetic island appears. The reconnection rate reaches above $0.035$ after the second magnetic island appears, resulting in five-fold increase in the reconnection rate in the lower resolution run of Case~III during a very short period. Though the highest reconnection rate in this work and that in \cite{2013PhPl...20f1202L} are close, the physical mechanisms for increasing the reconnection rates are very different in the two studies. In \cite{2013PhPl...20f1202L}, the main mechanism for increasing the reconnection rate is the strong recombination effect. The plasmoid instabilities in their simulations only increase the reconnection rate by approximately $15\%$. In this work, the strong ionization rate in our low $\beta$ simulations makes the neutral-ion collisional mean free path within the current sheet to be much smaller than the width of the current sheet, and the ionized and neutral fluid flows are well-coupled throughout the reconnection region. Consequently, the ionization-recombination non-equilibrium effects do not lead to fast magnetic reconnection in this regime. The plasmoid instability is still the main mechanism to lead the fast magnetic reconnection in Case~III in this work. As mentioned in the previous paragraph, the higher resolution simulation of Case~III is terminated when the plasmoid instability just develops into the nonlinear stage, the sharp increase of the reconnection rate also does not appear yet. However, one can find that the reconnection rate sharply increases to a similar high value in both the higher and the lower resolution runs of Case~II. Therefore, we can expect that the reconnection rate in the higher resolution run of Case~III will also behave similarly as that in the lower resolution run. Yet higher resolution simulations of Case III will be pursued in future work to track multi-scale plasmoid evolution further into the non-linear regime.  

As shown in Figure~1 and 3, the plasmoid instabilities terminate at the reconnection current sheet length of $100$~m and width of $2$~m. The plasmas inside the current sheet become strongly ionized during the reconnection process and the ion density can reach about $0.5\times10^{21}$~m$^{-3}$. Then, inside the current sheet the neutral-ion collisional mean free path and the ion inertial length are $\lambda_{ni} \approx 2.37\times 10^{-3}$~m and $d_{i} \approx 0.01$~m, which are both much smaller than the width of the thinnest current sheet. Therefore, collisionless effects can also be ignored within reconnection current sheets in strongly magnetized regions around the solar TMR. Though the recombination effect does not significantly increase the reconnection rate in such strongly magnetized regions, the non-equilibrium ionization-recombination effect is still important. As discussed in  \citep{2018ApJ...852...95N}, the non-equilibrium ionization-recombination factor leads to much lower temperature increases. 

By comparing the numerical results in Case~A, I, II and III, we find that the ionized and neutral fluid flows are well-coupled throughout the reconnection region even after plasmoid instabilities appear in all the cases (not shown). The ionization rate is also higher than the recombination rate in the reconnection process, and most of the generated thermal energy is also radiated away through radiative cooling in all the cases (not shown).

 
 \subsection{Magnetic reconnection at different heights in the low solar atmosphere}\label{ss:MR:II}
 The plasma density in Case~IV is 25 times higher than that in Case~I, and the plasma density in Case~V is 5 times lower than that in Case~I. According to the VAL-C solar atmosphere model \citep{1981ApJS...45..635V}, Cases~IV, I and V represent magnetic reconnection at $300$~km above the solar surface in the photosphere, $600$~km above the solar surface around the temperature minimum region and $800$~km above the solar surface in the low solar chromosphere, respectively. The characteristic length scale in Case~I, IV and V is the same with  $L_{\star1}=200$~m. The strength of reconnection magnetic field in Case~IV is $2500$~G, which is 5 times higher than that in Case~I and Case~V. The initial plasma $\beta$ in both Case~I and Case~IV is 0.058. 
 
Figure~4(a) shows the evolution of the calculated Lundquist number $S_{sim}$ with time before plasmoid instabilities appear, and Figure~4(b) shows reconnection rate $M_{sim}$ versus time before and after plasmoid instabilities appear. Magnetic reconnection starts at about $t=14t_{\star1}$ in Case~I and Case~IV, and it starts at about $t=8.5t_{\star1}$ in Case~V. $t_{p1}$ corresponds to the time point when the first magnetic island starts to appear in Cases~I and V. The Lundquist number is about $3400$ just before the first magnetic island appears in Case~I. At the end of the simulation in Case~IV, the Lundquist number is about $2000$, which is too low to result in the plasmoid instabilities, and no magnetic islands appear in this case. After $t=17t_{\star1}$, the quick increase of the plasma temperature at the reconnection X-point in Case~V leads to the decrease of magnetic diffusion and the increase of Lundquist number $S_{sim}$, the value of $S_{sim}$ eventually reaches about $1.4\times10^4$ in Case V before the first magnetic island appears. As shown in Figure~4(b), magnetic reconnection basically follows the Sweet-Parker model with $M_{sim} \sim S_{sim}^{-1/2}$ before the plasmoid instabilities appear. After the first magnetic islands appear, the reconnection rate starts to increase with time in Cases~I and V.  From these results, we can also conclude that the plasmoid cascading process will terminate at a larger scale in an environment with higher plasma density or weaker reconnection magnetic fields.          
 
Figure~5 shows the distributions of current density $J_z/J_{\star1}$ and ionization fraction $f_i$ at three different time points in Cases~I, IV and V, respectively.  The maximum current density during the later stage of the reconnection process reaches the greatest value in Case~IV with the strongest reconnection magnetic fields. Though the strength of reconnection magnetic fields is the same in Case~I and V, the maximum current density in Case~V is about two times greater than that in Case~I. Therefore, the current density generated in the reconnection process relates to both the strength of reconnection magnetic fields and plasma density. The stronger reconnection magnetic fields and lower plasma density both can result in the greater maximum current density in the reconnection process. In the case with higher plasma density, the temperature increase is lower and the magnetic diffusion is higher. The higher plasma density also results in the stronger viscous diffusion. Therefore, the current is more strongly dissipated by higher magnetic and viscous diffusion in the case with higher plasma density. However, the results in Figure~5 show that the current density depends on the strength of magnetic field more strongly than the plasma density. The initial values of plasma $\beta$ and ionization fraction in Case~I and Case~IV are the same, but the maximum ionization fraction in Case~IV is much smaller than that in Case~I during the later stages of the magnetic reconnection process. The maximum ionization fraction reaches the greatest value in Case V with the minimum initial plasma density. The stronger reconnection magnetic fields also result in the higher maximum ionization fraction \citep{2018ApJ...852...95N}. How high the maximum ionization fraction can reach during a reconnection process also depends on both the initial plasma density and the strength of reconnection magnetic fields. Figure~5 also shows that the width of the current sheet during the later reconnection stage is minimum in Case~V, which is measured about $1.5$~m by using the half-width at the half-maximum in the current density $J_z/J_{\star1}$. Though the plasma density in Case~V is five times lower than that in Case~I, we find that the width of the current sheet in Case~V is still more than one order of magnitude larger than both the ion-neutral collisional mean free path $\lambda_{in}$ and the ion inertial length $d_i$ during the magnetic reconnection process.  

Figure~6 shows the profiles of ion and neutral temperatures across the reconnection X-point along y-direction in Cases~I, IV and V at the same three time instances for each of the Cases as in Figure~5. The plasma temperatures inside the current sheet increase with time during the reconnection process in each of the three cases. The maximum temperatures in Cases~I, IV and V reach $1.8\times10^4$~K, $1.3\times10^4$~K and $3.1\times10^4$~K before the simulations are terminated due to lack of spatial resolution. The simulation results show that it is very difficult to heat the much denser photosphere plasmas to high temperatures. The plasma temperature over $2\times10^4$~K is not observed in Case~IV even when the reconnection magnetic field is as strong as $2500$~G. Though, we cannot discard the possibility that the plasma temperature could exceed $2\times 10^4$~K in Case IV if the simulation could run for a significantly longer period of time.  In Case~V with much lower number density, the plasma can be easily heated above $3\times10^4$~K as shown in our simulations. 

\section{Summary and discussions}\label{s:discussion}
In this work, we have used the reactive multi-fluid plasma-neutral module of the HiFi modeling framework to study magnetic reconnection in initially weakly ionized plasmas around the solar TMR on different spatial scales and also at three different heights from $300$~km to $800$~km above the solar surface. The main conclusions are as follows:

(1) In strongly magnetized solar TMR, the critical Lundquist number for the onset of the plasmoid instability is between $2000\sim3400$ and the corresponding critical aspect ratio is between $45\sim70$ in current sheets with guide field evolving from a force-free magnetic configuration in 2.5D simulations. These critical values are about five times smaller than those in the previous work by \cite{2012ApJ...760..109L, 2013PhPl...20f1202L}, which considered magnetic reconnection in weakly magnetized middle chromosphere in a Harris Sheet resulting in null-point reconnection. The different plasma $\beta$ and configurations of the reconnecting magnetic field are the main reasons for this substantial difference. The critical Lundquist number and aspect ratio in this work are very close to the results in previous work by \cite{2013PhPl...20f1206N}, in which magnetic reconnection in fully ionized plasmas were studied by using the same initial force-free magnetic field configuration.. 

(2) The numerical results of magnetic reconnection at different scales show that the plasmoid cascading process terminates at a length scale of $100$~m and width of $2$~m in the strongly magnetized solar TMR with reconnection magnetic fields of $500$~G and plasma $\beta$ of 0.058. These length scales are much larger than the ion inertial length and the ion-neutral collisional mean free path within the resulting current sheets, the collisionless effects and the decoupling effects of ions and neutrals on magnetic reconnection rate are small. The plasmoid cascading process terminates at a larger scale in an environment with higher plasma density and weaker reconnection magnetic fields. 

(3)  In strongly magnetized solar TMR region, the plasmoid instability plays an important role in accelerating magnetic reconnection. The reconnection rate scales with the Lunqusit number as $M_{sim} \sim S_{sim}^{-1/2}$ before plasmoid instabilities appear. After plasmoid instabilities appear, the reconnection rate sharply increases to a high value (about 0.035), which is independent of the Lundquist number. These characteristics are very similar to magnetic reconnection in fully ionized plasmas. The recombination effect does not significantly affect the reconnection rate in this region, which is very different from the previous weakly ionized higher plasma $\beta$ simulations \citep[e.g.,][]{2012ApJ...760..109L, 2013PhPl...20f1202L, 2017ApJ...842..117A}. 

(4) A weakly ionized low chromosphere plasmas with the neutral density of $10^{20}$~m$^{-3}$ can be heated from several thousand Kelvin to above $3\times10^4$~K during a magnetic reconnection process with magnetic fields of $500$~G.  We find that the much denser photosphere plasmas with the neutral density of $1.25\times10^{22}$~m$^{-3}$ are heated to only about $1.3\times10^4$~K with strong reconnection magnetic fields of $2500$~G.

(5) A stronger reconnection magnetic fields and lower plasma density can both result in the higher maximum current density and ionization fraction in a reconnection process with initially weakly ionized plasmas. However, the maximum current density depends on the strength of magnetic field more strongly than the plasma number density. 

The numerical results in this work indicate that the characteristics of guide field magnetic reconnection in strongly magnetized low solar atmosphere ($300-800$~km above the solar surface) are close to that of the fully ionized plasmas. The initially weakly ionized dense plasmas in the current sheet eventually become strongly ionized during the reconnection process. Then, the high ion density makes the ion-neutral collisional mean free path to be much smaller than the width of the current sheet, then the decoupling of ionized and neutral flows is not obvious and the significant increase of magnetic reconnection rate by recombination effect as shown in the previous weakly magnetized high plasma $\beta$ simulations does not appear. The length scales of the observed EBs and IBs are usually about $10^6$~m, which is two orders of magnitude larger than the characteristic length scale of $L_{\star3}=10^4$~m in Case~III. Therefore, the Lundquist number in these reconnection events are above $10^6$, which is much higher than the critical Lundquist number predicted from this work. We conclude that the plasmoid instability instead of the recombination effect is the dominated mechanism to result in fast magnetic reconnection in strongly magnetized regions around the solar TMR. Though the ionized and neutral fluids are coupled well in the reconnection site, it does not mean that the one-fluid model is good enough to study magnetic reconnection in this region.  \cite{2018ApJ...852...95N} has shown the ionization-recombination non-equilibrium dynamics are very important for studying the evolutions of the ionization fraction and plasma temperatures in the reconnection process. Therefore, the reactive multi-fluid plasma-neutral model applied in this work  is likely necessary for studying magnetic reconnection in this region. Since the width of the current sheet is also much larger than the ion inertial length inside the current sheet during the reconnection process, the collisionless effects on magnetic reconnection are negligible in this strongly magnetized region. 

The simulations in this work were terminated not long after the simulations enter the plasmoid regime. Since the plasma temperatures inside the current sheet increase with time during the reconnection process, one may expect that the plasmas can be heated to higher temperatures if the simulations can last for a longer period. In future work, we will aim to extend the large-scale simulations further into the cascading plasmoid regime to study plasma heating during the saturation phase of the multi-scale current sheet evolution.

\section*{ACKNOWLEDGMENTS}
 This research is supported by NSFC Grants 11573064, 11333007, 11303101 and 11403100; the Western Light of Chinese Academy of Sciences 2014; the Youth Innovation Promotion Association CAS 2017; the Applied Basic Research of Yunnan Province in China Grant 2018FB009;  the key Laboratory of Solar Activity grant KLSA201812; Program 973 grant 2013CBA01503; NSFC-CAS Joint Grant U1631130; CAS grant QYZDJ-SSW-SLH012; the Special Program for Applied Research on Super Computation of the NSFC-Guangdong Joint Fund (nsfc2015-460, nsfc2015-463, the second phase); and the Strategic Priority Research Program of CAS with grant XDA17040507. The authors gratefully acknowledge the computing time granted by the Yunnan Astronomical Observatories and the National Supercomputer Center in Guangzhou, and provided on the facilities at the Supercomputing Platform, as well as the help from all faculties of the Platform. V.S.L. acknowledges support from the US National Science Foundation (NSF).

\clearpage

\begin{table}
 \caption{The different characteristic length scales $L_{\star}$ and times $t_{\star}$, and the initial parameters in the five cases. The characteristic plasma number density $n_{\star}=10^{21}$~m$^{-3}$, the characteristic magnetic field $B_{\star}=0.05$~T$=500$~G , the characteristic velocity $V_{\star}=34.613$ km\,s$^{-1}$ and the characteristic temperature $T_{\star}=1.441\times10^5$~K are the same in all the five cases. }
 \label{Parameter}
    \begin{tabular}{lcccccccc}
     \hline
                            & $L_{\star}$ (m) & $t_{\star}$ (s) & $n_{n0}$ & $f_{i0}$ & $T_{i0}$ & $b_{p}$ & $\beta_0$ & Radiative cooling \\
      \hline            
         Case A        &    $100$  & 0.0029  & 0.5 $n_{\star}$  & $10^{-4}$  & $0.05829 T_{\star}$ & 1  & 0.058 & $L_{rad1}$  \\                        
      \hline
         Case I        &    $200$  & 0.0058  & 0.5 $n_{\star}$  & $10^{-4}$  & $0.05829 T_{\star}$ & 1  & 0.058 & $L_{rad1}$ \\ 
       \hline  
         Case II       &    $10^3$ & 0.029  & 0.5 $n_{\star}$  & $10^{-4}$  &  $0.05829 T_{\star}$ &1   & 0.058  &  $L_{rad1}$ \\
       \hline
          Case III      &    $10^4$ & 0.29  &  0.5 $n_{\star}$  & $10^{-4}$  &  $0.05829 T_{\star}$ &1   & 0.058   & $L_{rad1}$ \\
       \hline
          Case IV   &     $200$ & 0.0058  &  12.5 $n_{\star}$  & $10^{-4}$  &  $0.05829 T_{\star}$ & 5   & 0.058  & $L_{rad1}$  \\
       \hline
          Case V   &     $200$ & 0.0058  &  0.1 $n_{\star}$  & $10^{-3}$  &  $0.06559 T_{\star}$ & 1   & 0.01305  &  $L_{rad2}$ \\     
       \hline       
   \end{tabular}
\end{table}

\begin{table}
 \caption{The smallest normalized grid sizes along the $y$-direction near $y=0$ in both the higher and lower resolution runs of Case~I, II, III, IV and V. }
 \label{Parameter}
   \begin{tabular}{lcccccc}
     \hline
                                                      &  Case~I                         & Case~II & Case~III & Case~IV & Case~V  \\
      \hline            
        lower resolution &$3.095\times10^{-4}$&$1.284\times10^{-4}$&$7.950\times10^{-5}$&$3.095\times10^{-4}$&$3.095\times10^{-4}$    \\                        
      \hline
        higher resolution&$1.855\times10^{-4}$&$3.309\times10^{-5}$&$3.309\times10^{-5}$&$1.855\times10^{-4}$&$1.855\times10^{-4}$   \\ 
       \hline  
    \end{tabular}
\end{table}

 \begin{figure}
      \centerline{\includegraphics[width=0.6\textwidth, clip=]{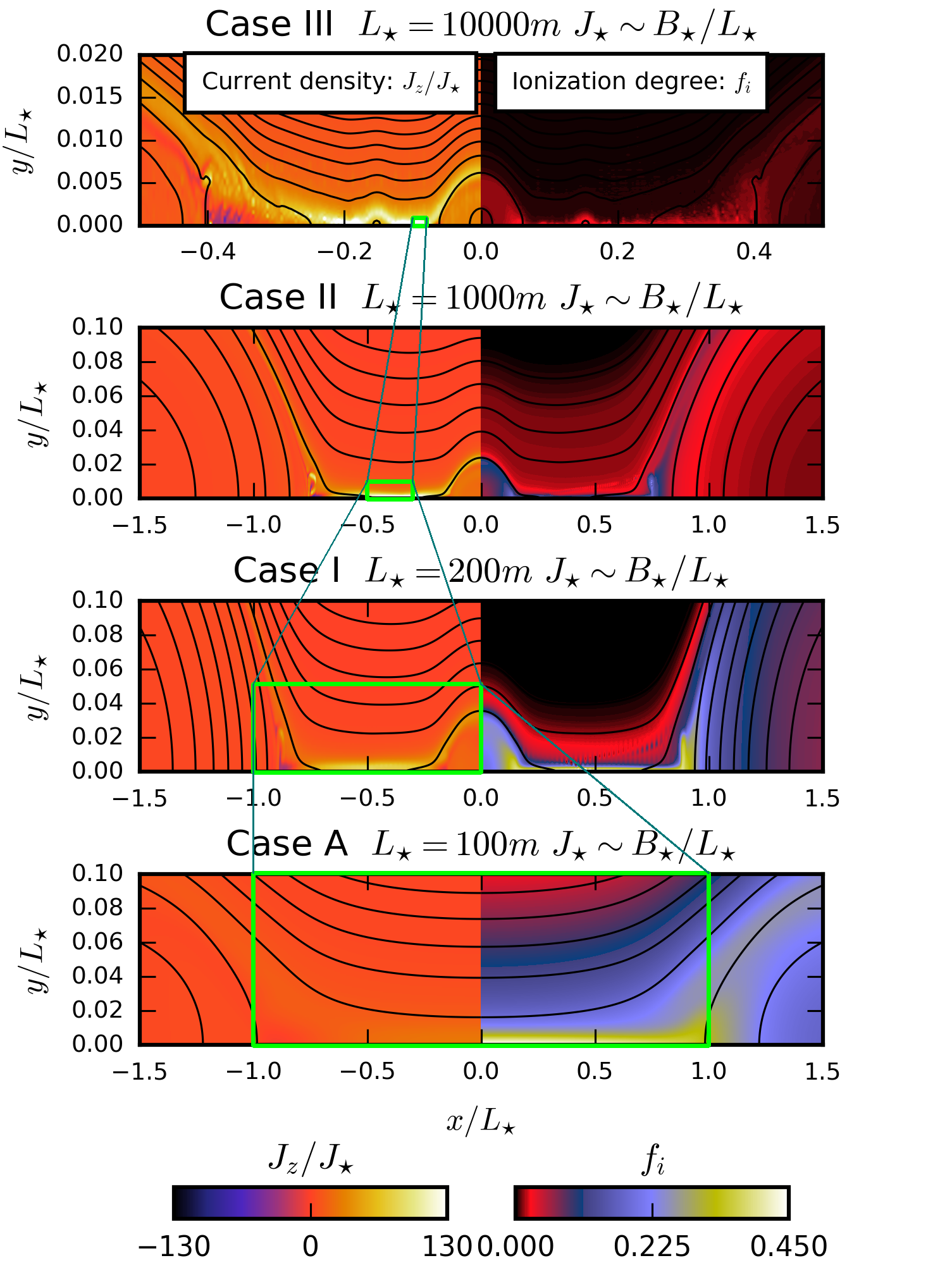} }
     \caption{Magnetic reconnection during the later stage at different length scales. The left side shows the distributions of the dimensionless current density $J_z/J_{\star1}$ and the right side shows the ionization fraction $f_i$ . The values of the characteristic length scale $L_{\star}$ are corresponding to $L_{\star1}=200$~m in Case~I, $L_{\star2}=1000$~m in Case~II and $L_{\star3}=10^4$~m in Case~III, respectively. One should note that the values of the characteristic current density $J_{\star}$ are also different in each case because of the different length scales, but the same color bar is used for all the cases.  Since the length scale in Case~A is smallest, the value of the characteristic current density $J_{\star}$ in Case~A is greatest, which makes the distribution of the normalized current density $J_z/J_{\star}$ in Case~A  be hardly visible. The green boxes correspond to the same size in each case with different length scales. }
    \label{fig.1}
 \end{figure}

\begin{figure}
      \centerline{\includegraphics[width=0.45\textwidth, clip=]{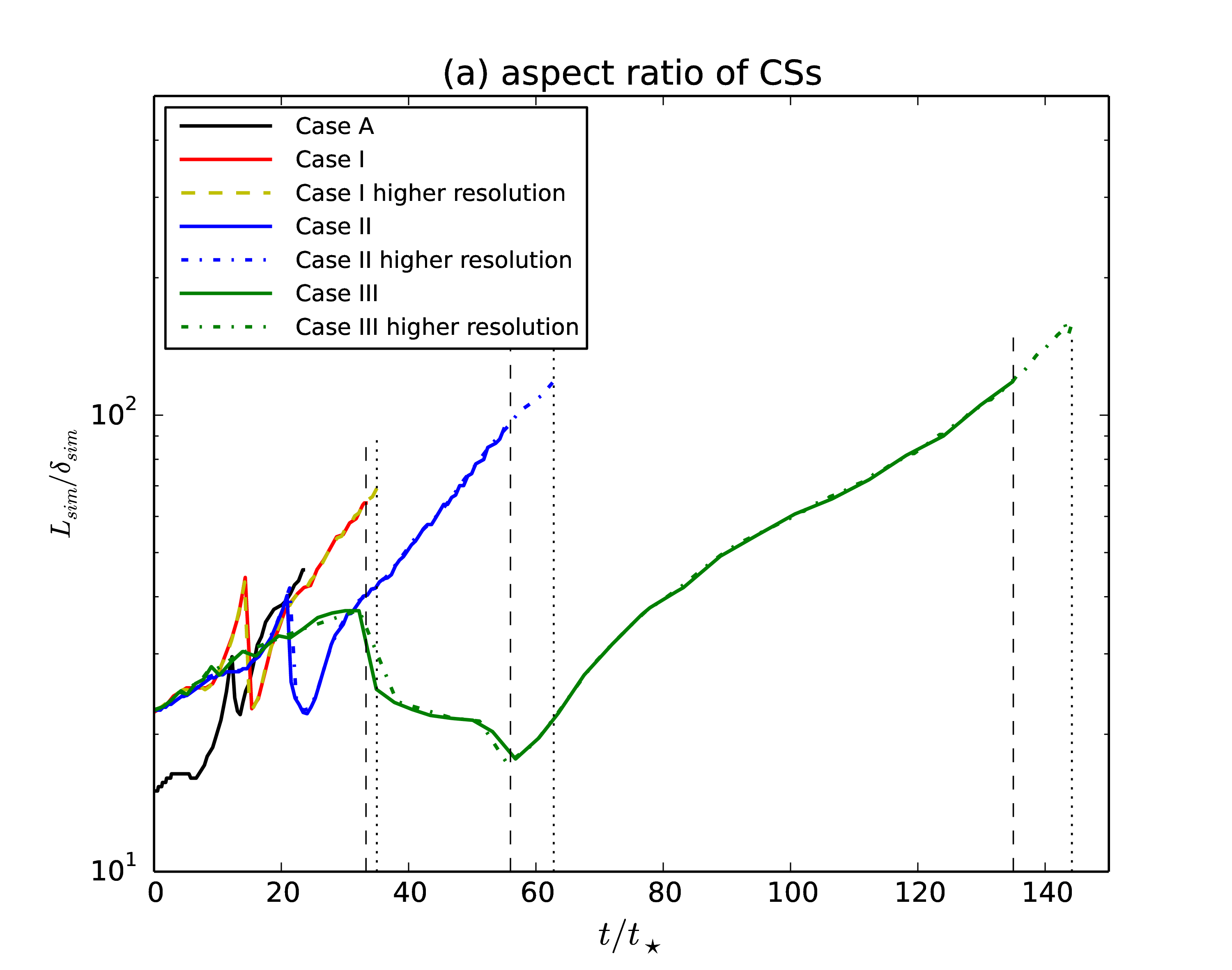}
                          \includegraphics[width=0.45\textwidth, clip=]{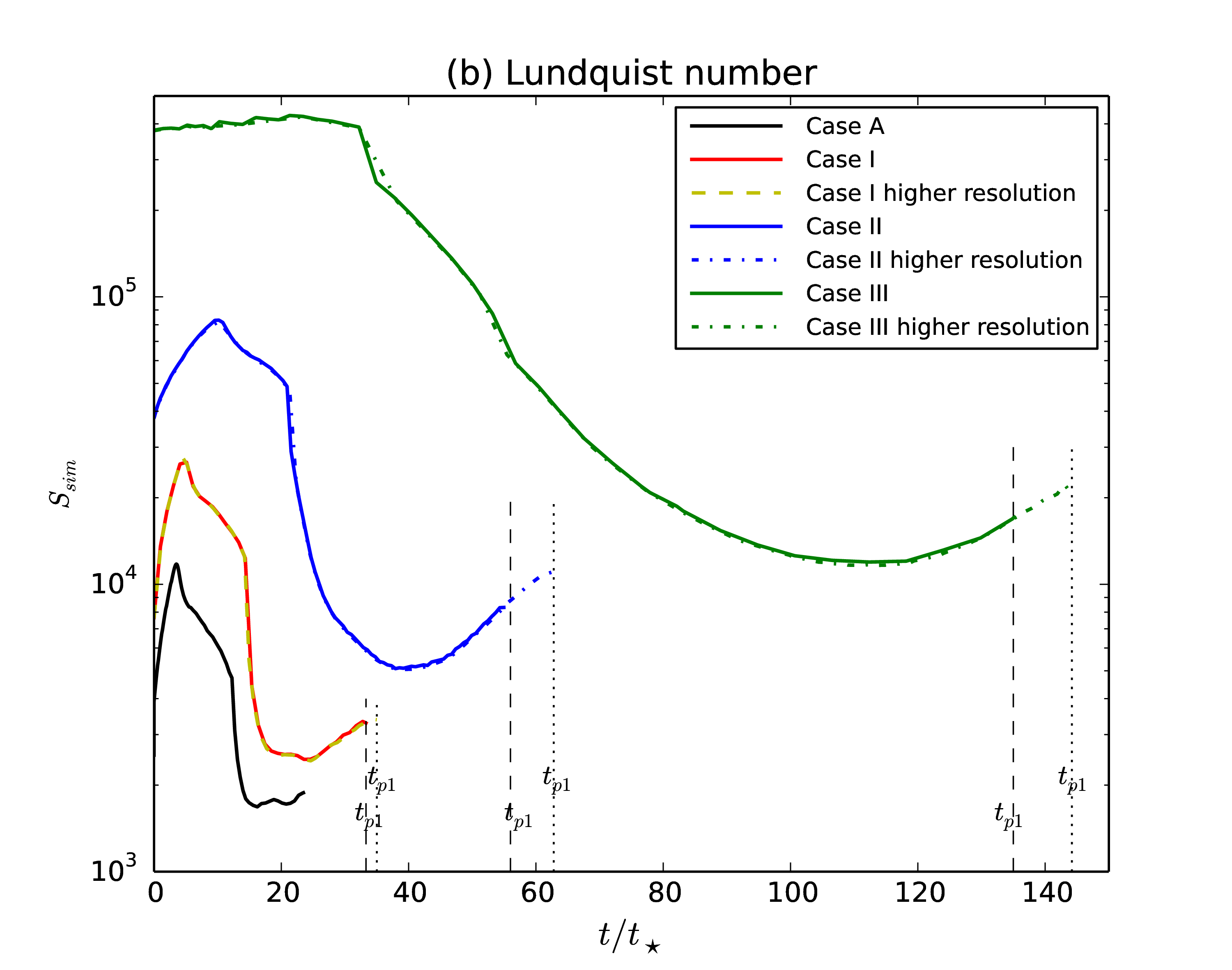} }
      \caption{(a) and (b) show the aspect ratio of the current sheets (CSs) $L_{sim}/\delta_{sim}$ and the calculated Lundquist number $S_{sim}$ versus time in Case~A in our previous work and in Cases~I, II and III in this work before the plasmoid instability appear. $t_{p1}$ represents the time point when the first magnetic island appears inside the current sheet in each run. The values of $t_{\star}$ are $t_{\star1}=0.0058$~s in Case~I, $t_{\star2}=0.029$~s in Case~II and $t_{\star3}=0.29$~s in Case~III, respectively.}
    \label{fig.2}
 \end{figure}
 
 \begin{figure}
      \centerline{\includegraphics[width=0.45\textwidth, clip=]{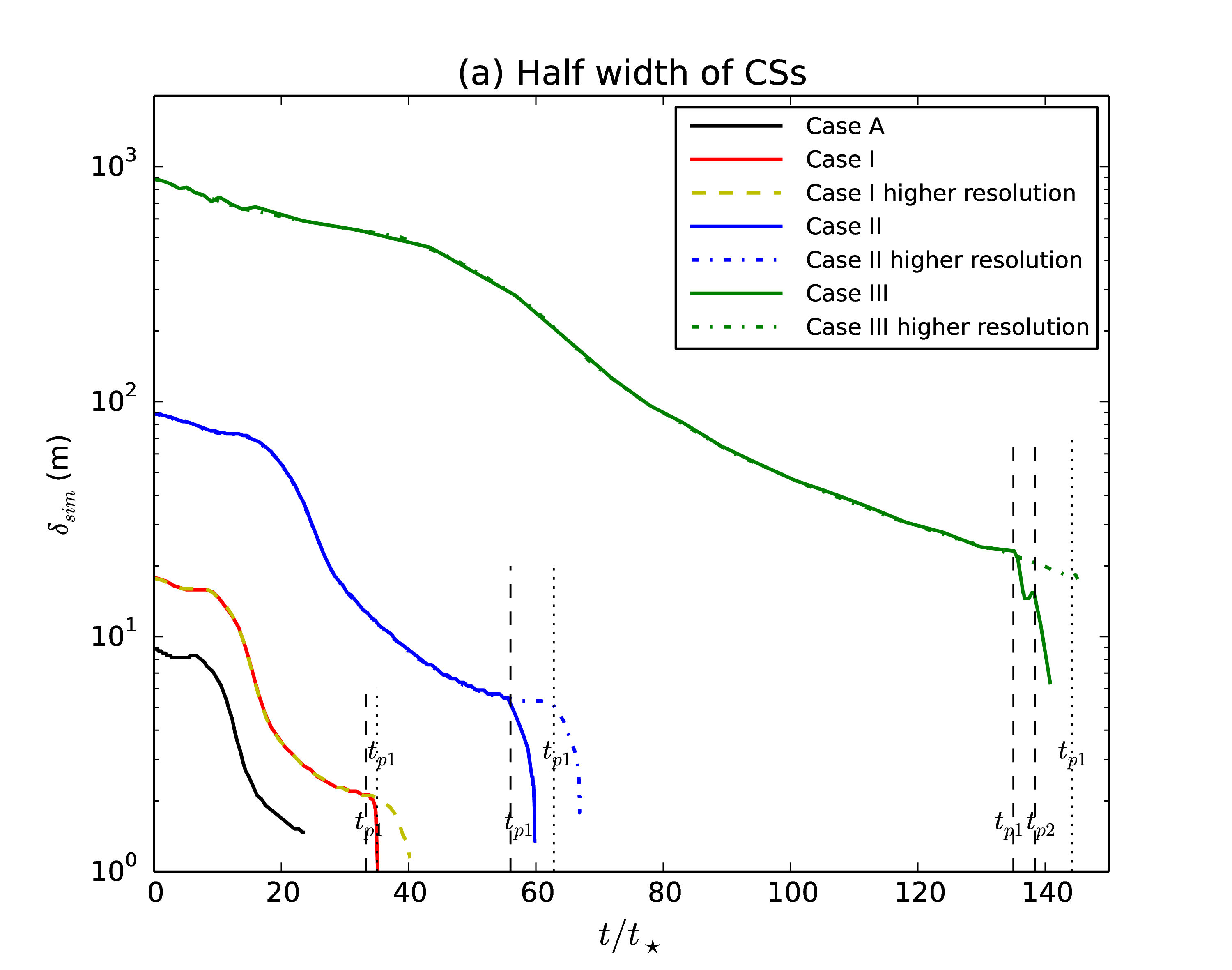}
                          \includegraphics[width=0.45\textwidth, clip=]{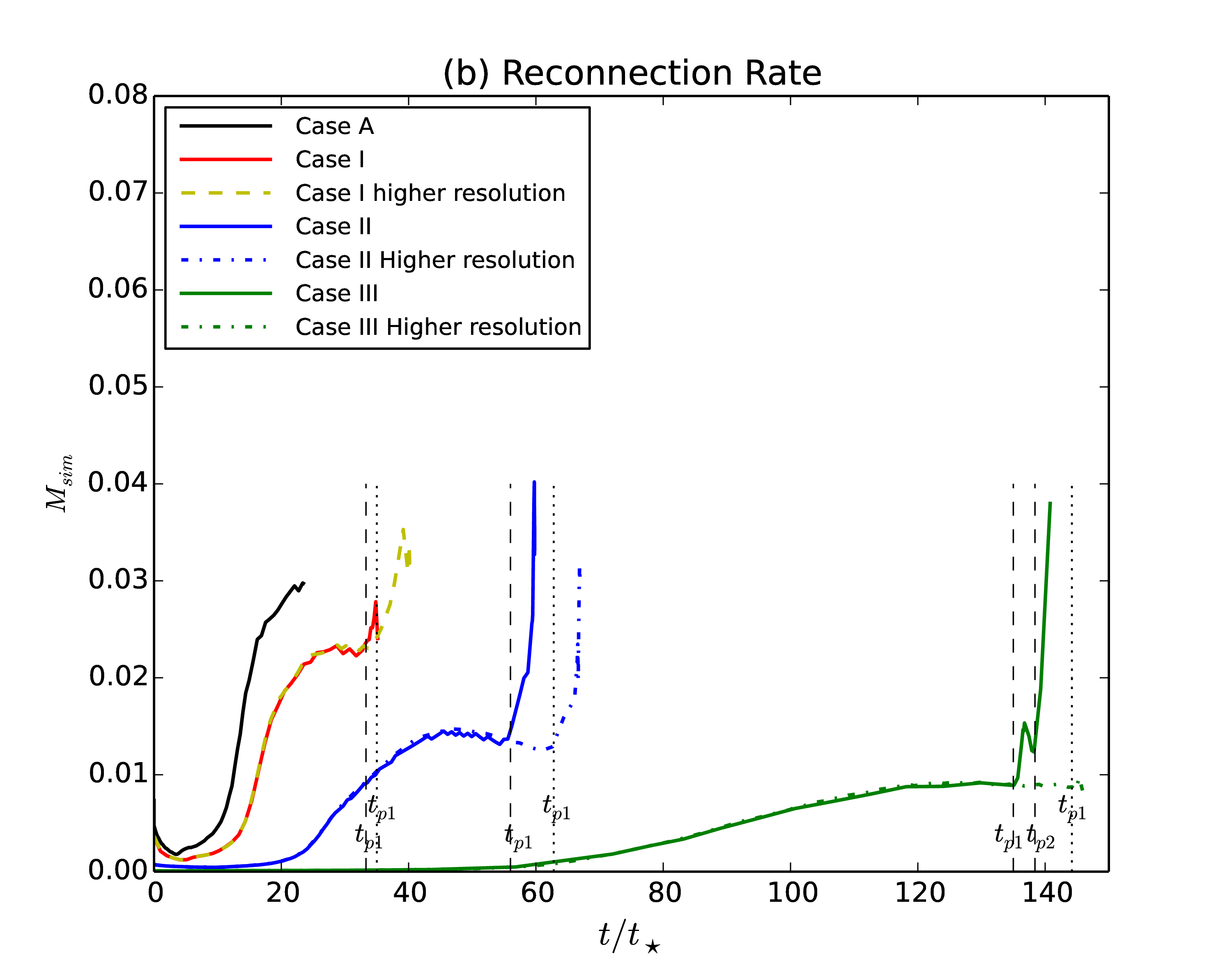} }
      \caption{(a) and (b) show the width of the current sheet $\delta_{sim}$ and  the calculated reconnection rate $M_{sim}$ versus time  in Case~A in our previous work and in Cases~I, II and III in this work before and after the plasmoid instability appear.  $t_{p1}$ represents the time point when the first magnetic island appear inside the current sheet in each run, $t_{p2}$ represents the time point when the second magnetic island appear inside the current sheet in the lower resolution run of Case~III. The values of $t_{\star}$ are $t_{\star1}=0.0058$~s in Case~I, $t_{\star2}=0.029$~s in Case~II and $t_{\star3}=0.29$~s in Case~III, respectively.}
    \label{fig.3}
 \end{figure}
 
  \begin{figure}
      \centerline{\includegraphics[width=0.45\textwidth, clip=]{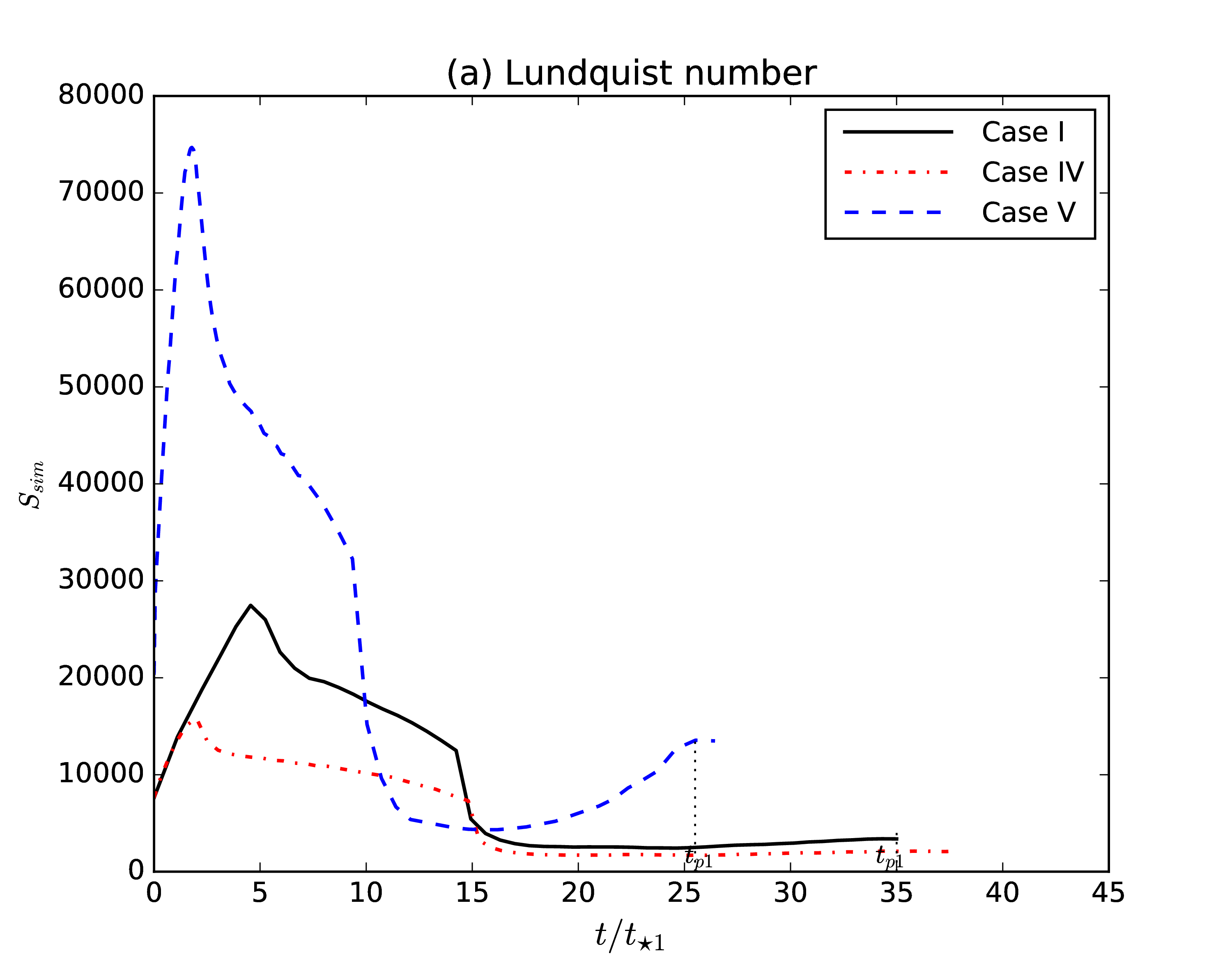}
                          \includegraphics[width=0.45\textwidth, clip=]{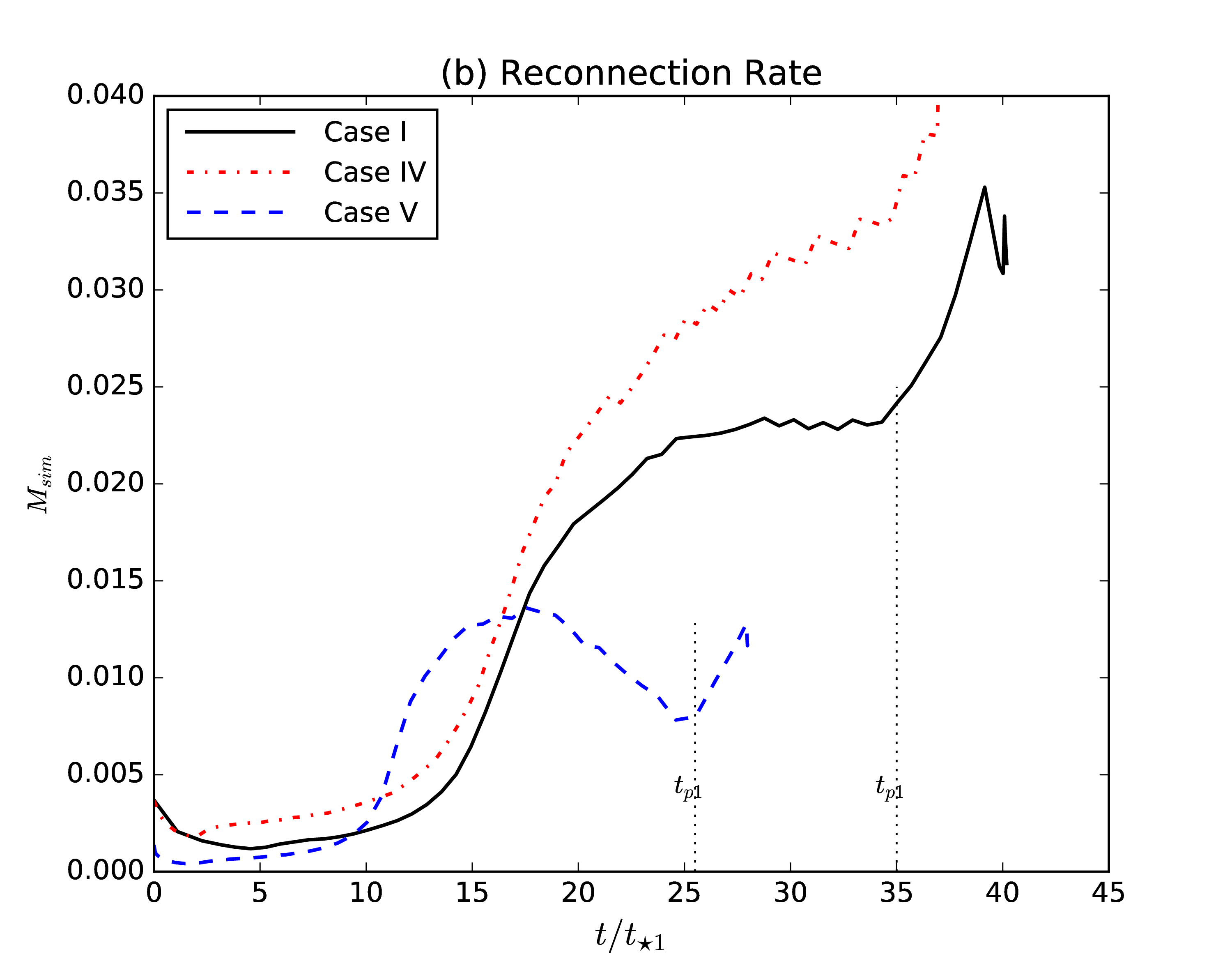} }
      \caption{(a) and (b) show the calculated Lundquist number $S_{sim}$ and reconnection rate $M_{sim}$ versus time  in Cases~I, IV and V.  $t_{p1}$ represents the time point when the first magnetic island appear inside the current sheet in Cases~I and V.}
    \label{fig.4}
 \end{figure}
 
 \begin{figure}
      \centerline{\includegraphics[width=0.33\textwidth, clip=]{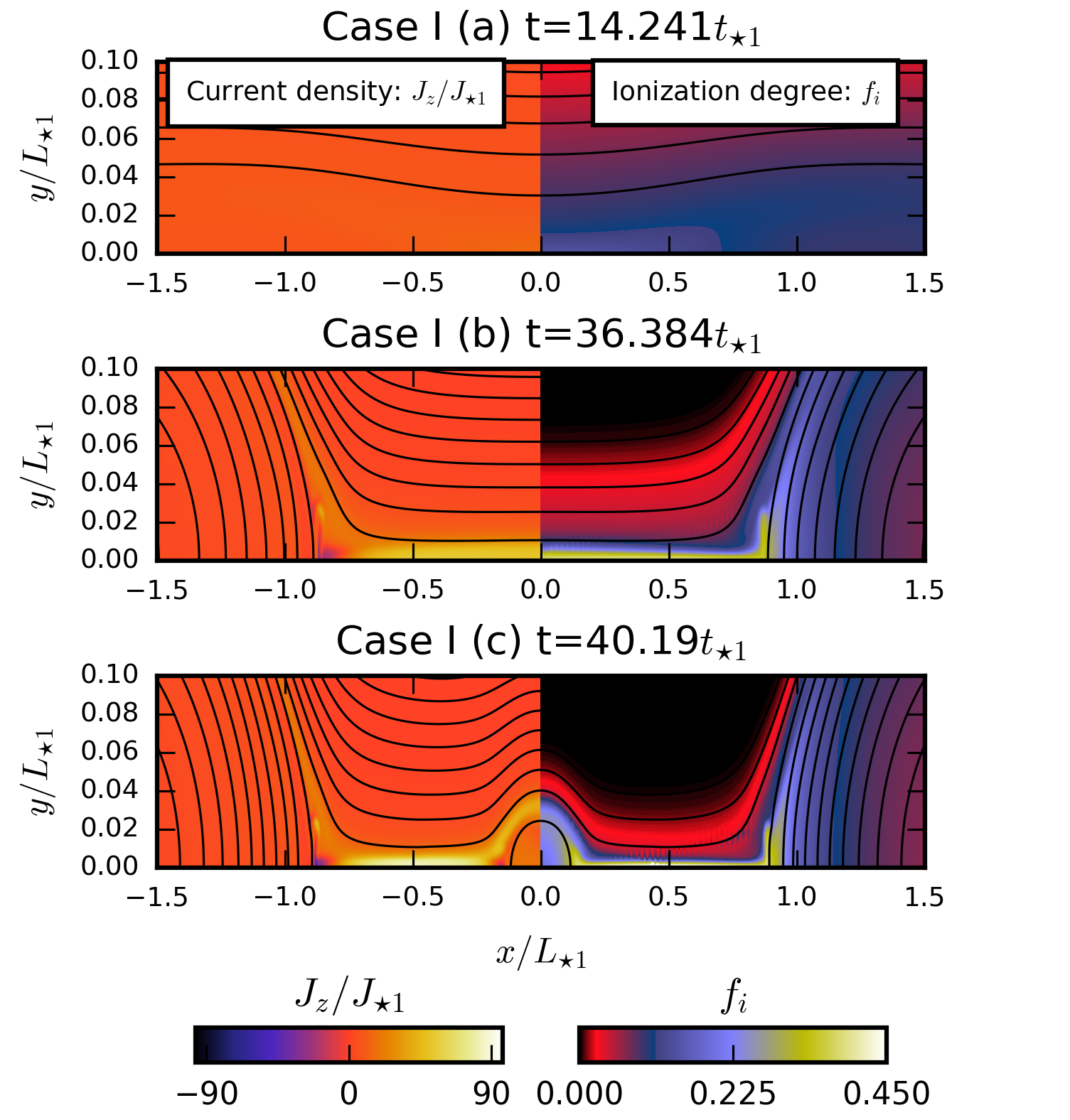}
                          \includegraphics[width=0.33\textwidth, clip=]{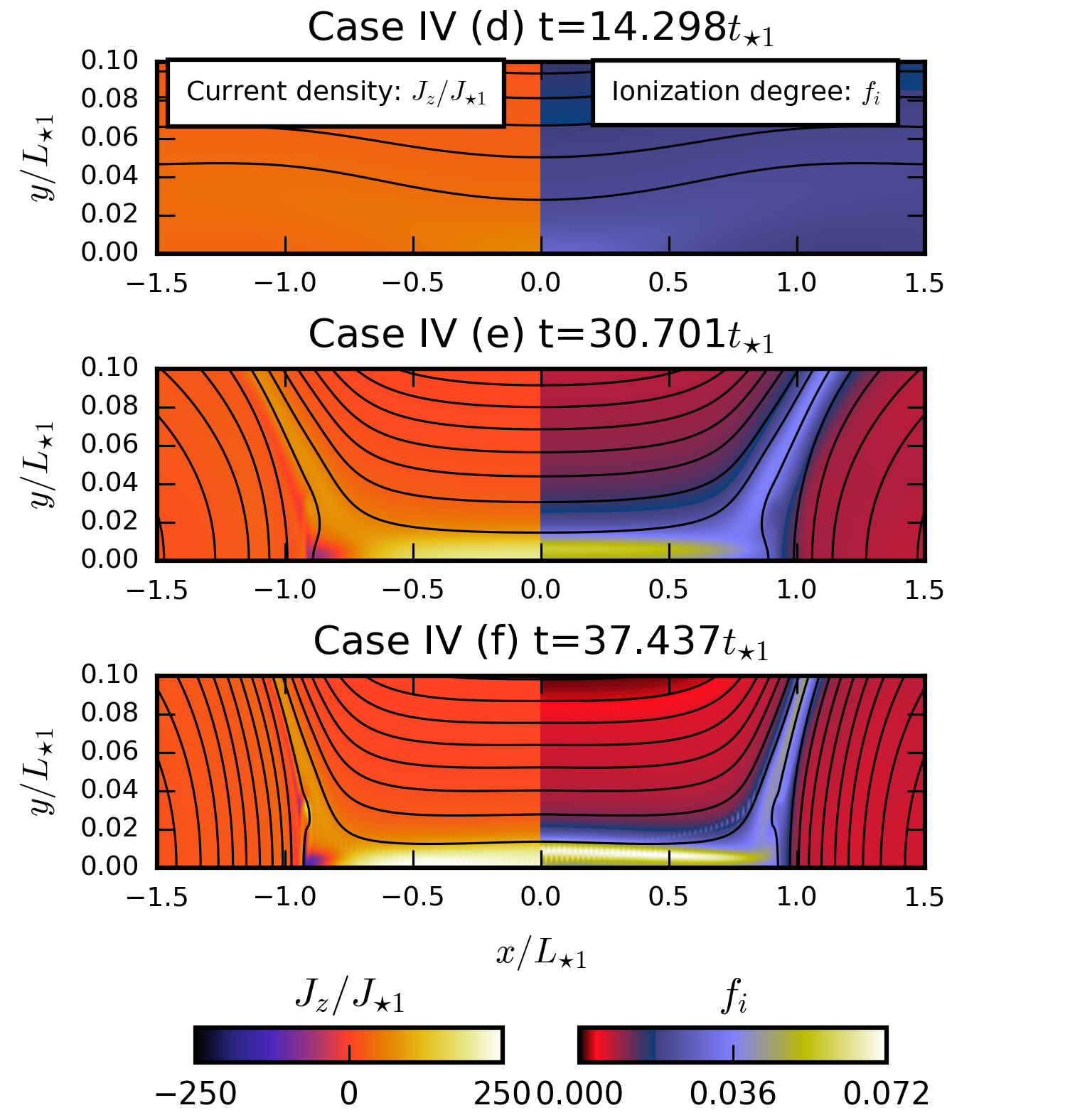} 
                           \includegraphics[width=0.33\textwidth, clip=]{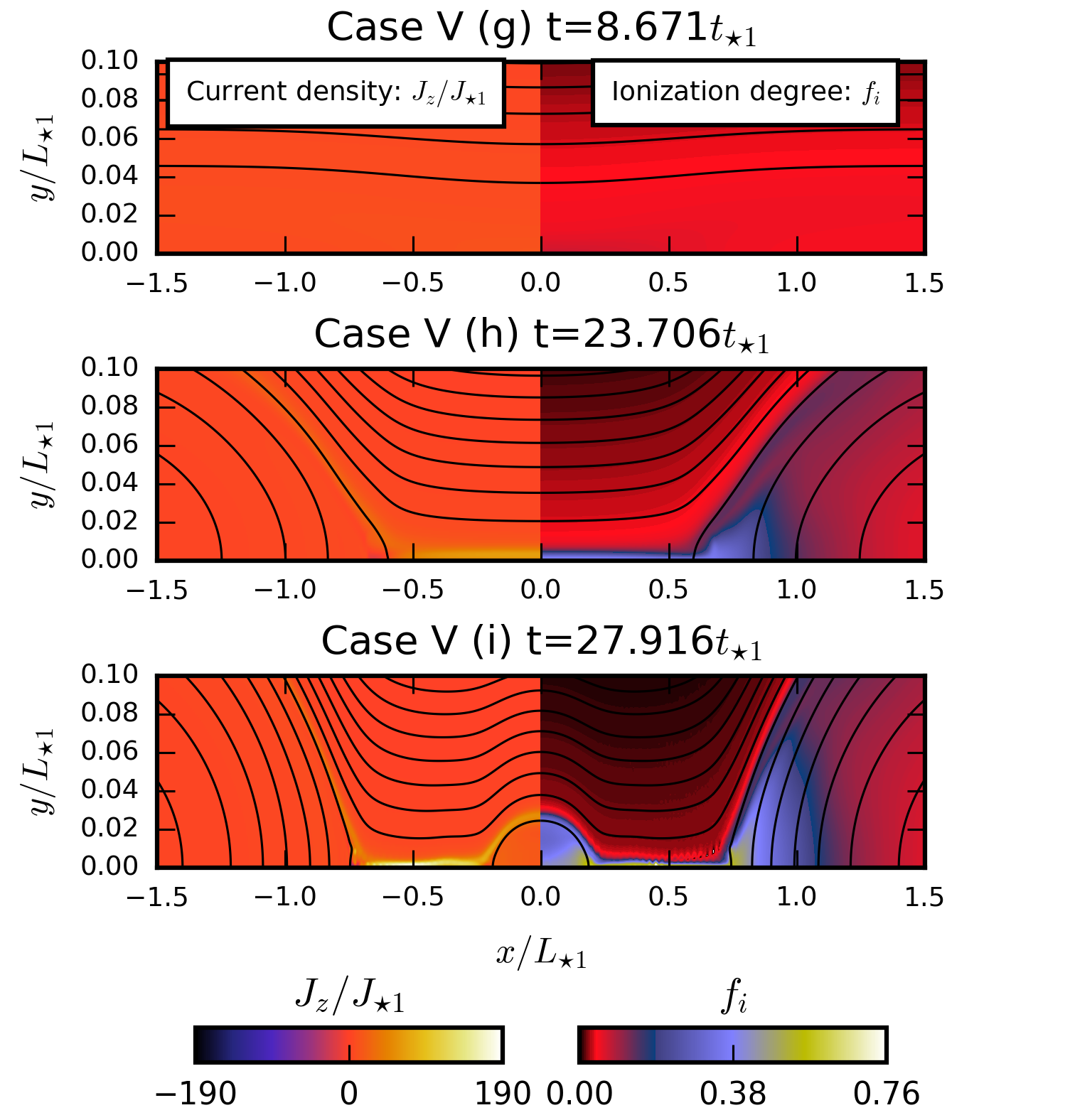} }
      \caption{The distributions of the dimensionless current density $J_z/J_{\star 1}$ (left) and the ionization fraction $f_i$ (right) in one quarter of the domain at three different times in Cases~I, IV and V. The current sheet lengths in Case~I at $t=14.241t_{\star1}$, in Case~IV at $t=14.298 t_{\star1}$ and in Case~V at $t=8.671 t_{\star1}$ are the same, as are those in (b), (e) and (h), and those in (c), (f) and (i).}
    \label{fig.5}
 \end{figure}
 
  \begin{figure}
      \centerline{\includegraphics[width=0.33\textwidth, clip=]{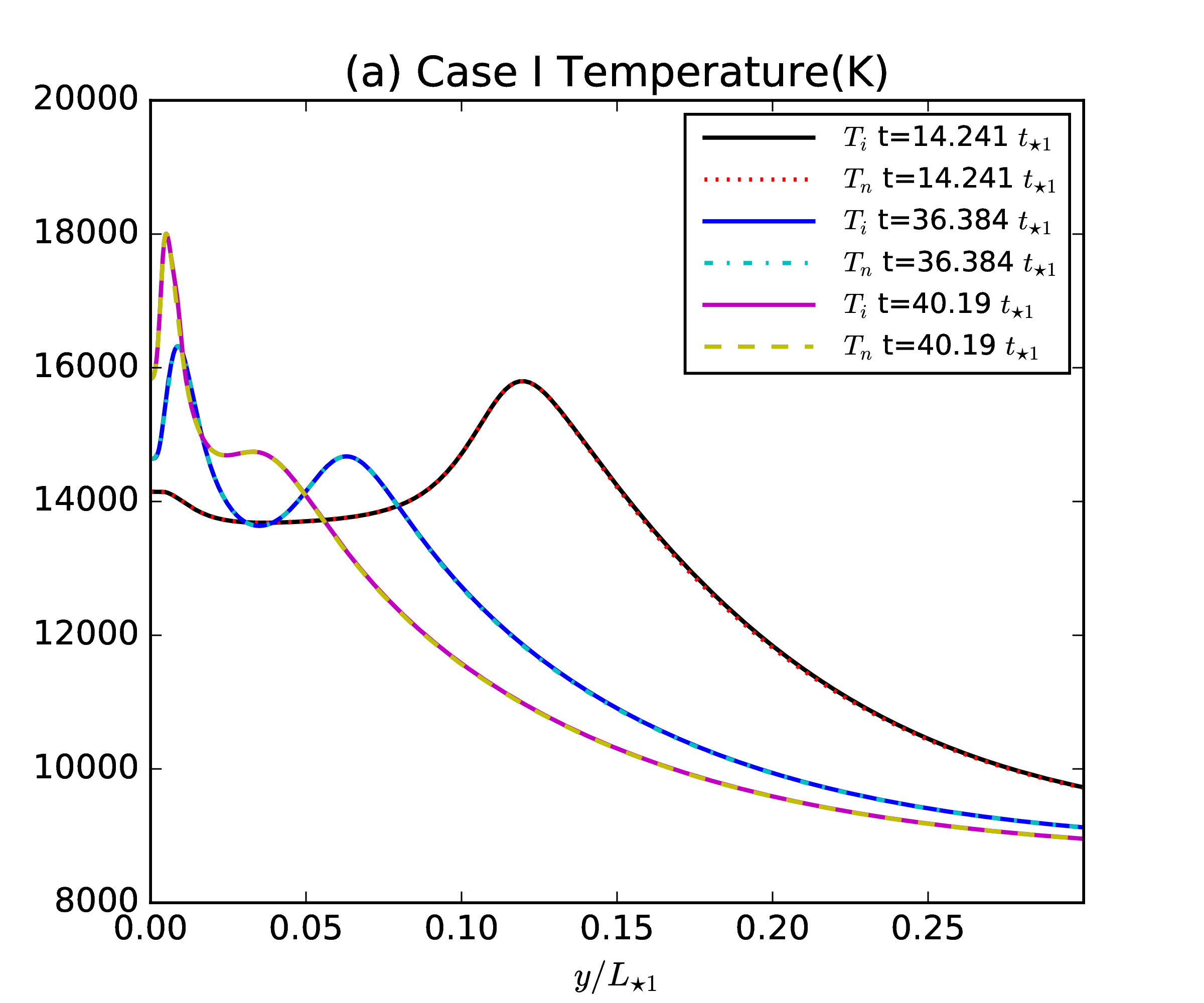}
                          \includegraphics[width=0.33\textwidth, clip=]{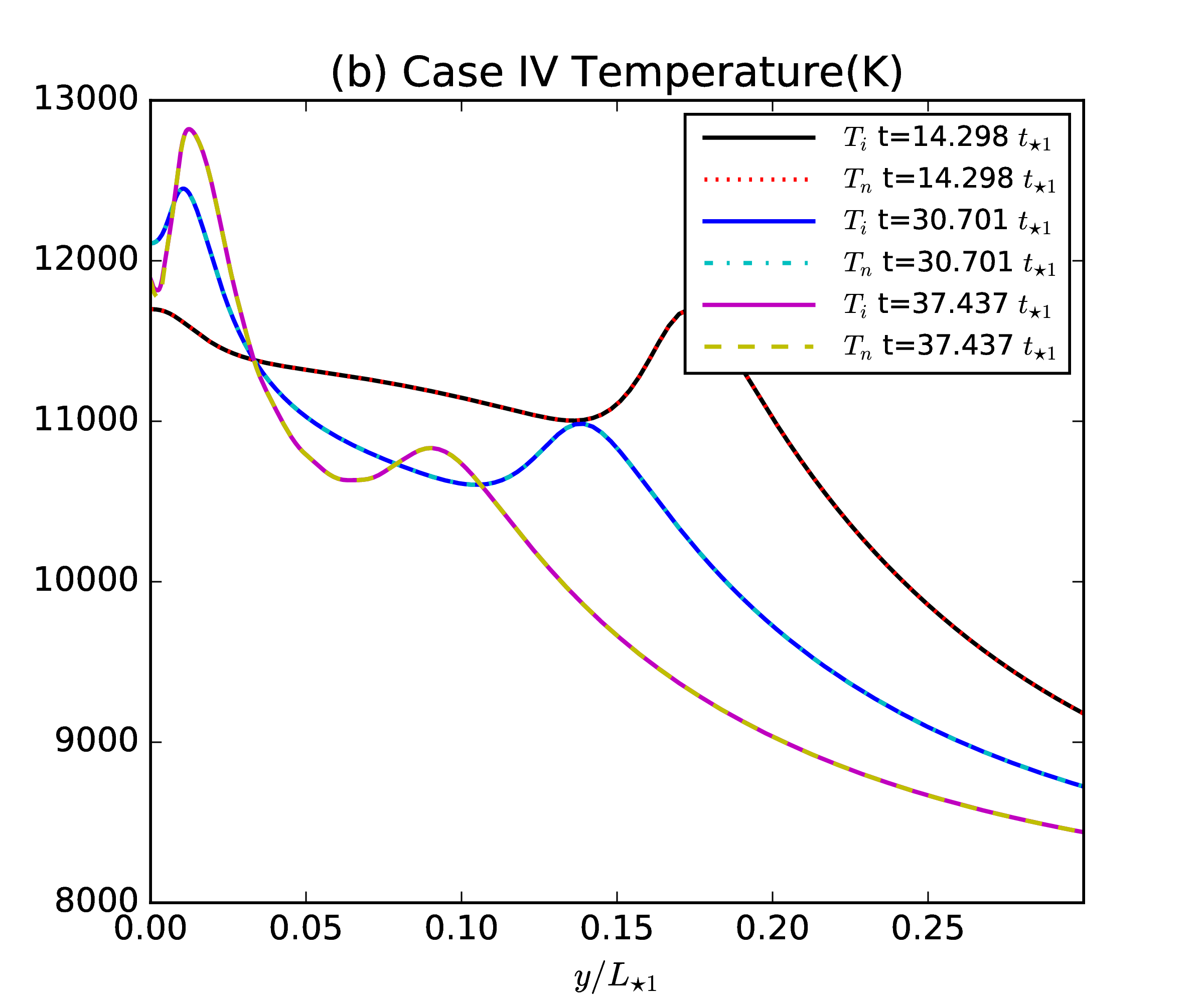} 
                           \includegraphics[width=0.33\textwidth, clip=]{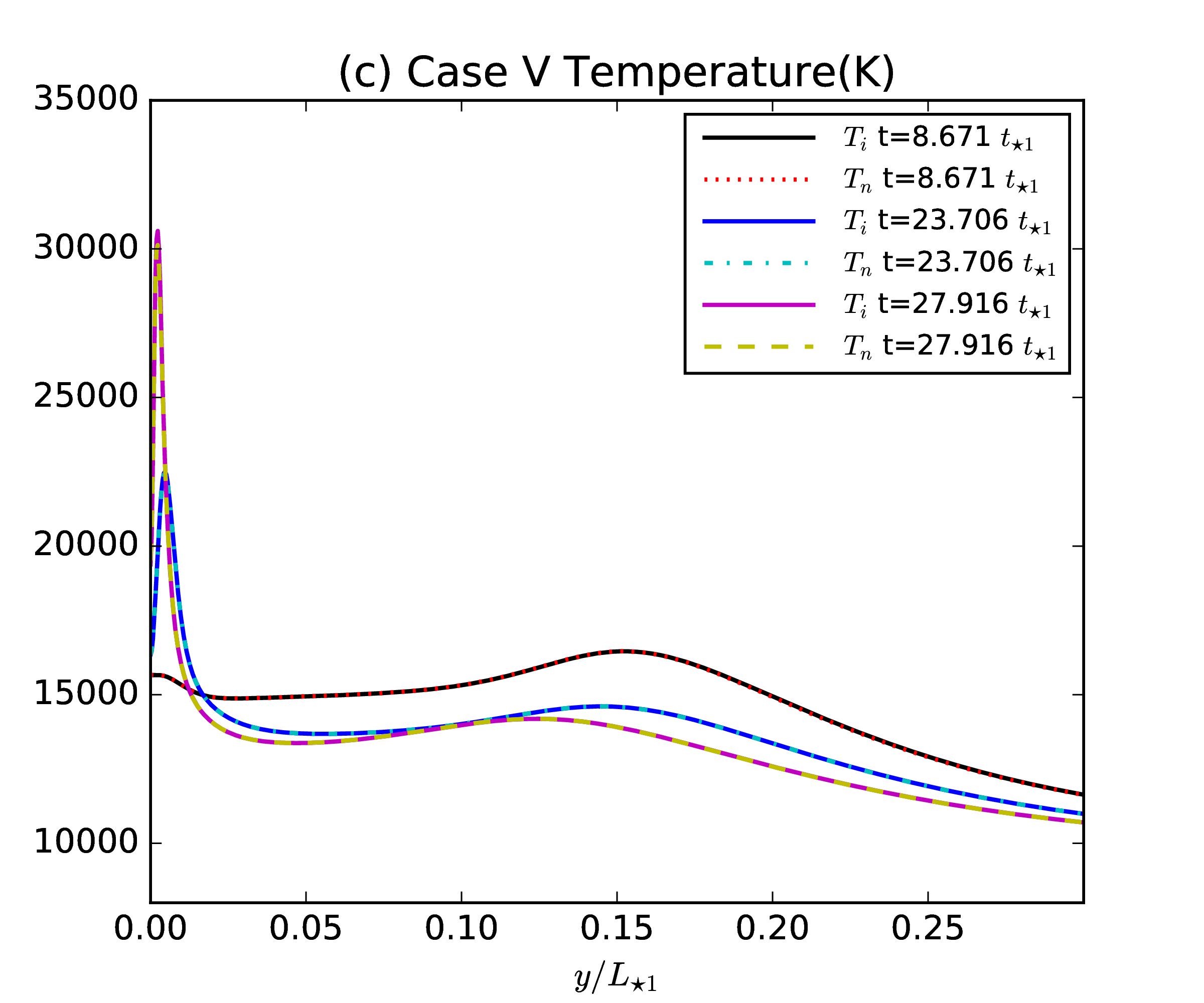} }
      \caption{The profiles of the ion temperature $T_{i}$ and neutral temperature $T_{n}$ in Kelvin across the reconnection X-point along  y-direction at the same times as shown as in Figure~5 in Cases~I, IV and V.}
    \label{fig.6}
 \end{figure}

---------------------------

\end{document}